\documentclass[preprint,aps,prd,numbers,sort&compress,merge]{revtex4-2}  % for double-spaced preprint

\usepackage{graphicx}  % needed for figures
\usepackage{bm}        % for math
\usepackage{amssymb}   % for math
\usepackage{slashed}   % for Dirac Slash
\usepackage{amsmath}   % for mutiline eqn
\usepackage{verbatim}  % for multi-line comment
\usepackage{color} % for textcolor
\usepackage{subfigure}
\usepackage[bookmarks=true,colorlinks=true,linkcolor=blue,unicode=true]{hyperref}

\begin{document}

\title{The study of neutral triple gauge couplings in the process \texorpdfstring{$e^+e^-\to Z\gamma$}{e+ e- to Z A} including unitarity bounds}

\author{Qing Fu}
\affiliation{Department of Physics, Liaoning Normal University, Dalian 116029, China}
\author{Ji-Chong Yang\email{yangjichong@lnnu.edu.cn}}
\affiliation{Department of Physics, Liaoning Normal University, Dalian 116029, China}
\author{Chong-Xing Yue\email{cxyue@lnnu.edu.cn}}
\affiliation{Department of Physics, Liaoning Normal University, Dalian 116029, China}
\author{Yu-Chen Guo\email{ycguo@lnnu.edu.cn}}
\affiliation{Department of Physics, Liaoning Normal University, Dalian 116029, China}

\begin{abstract}
The neutral triple gauge couplings~(nTGCs) provide a unique opportunity to probe new physics beyond the Standard Model.
The nTGCs can be described by an effective field theory~(EFT), which is valid only under a certain energy scale.
One of the signatures that an EFT is no longer valid is the violation of unitarity.
We study the partial wave unitarity bounds on the coefficients of nTGCs in the process $e^+e^-\to Z\gamma$.
In the experiments, the constraints obtained should be tighter than the unitarity bounds, otherwise the results are meaningless, therefore there exists a minimal luminosity for a $e^+e^-$ collider such as the CEPC to study the nTGCs.
To derive the minimal luminosity, the kinematic features and event selection strategy are studied by Monte-Carlo simulation.
Both the processes $e^+e^-\to \ell^+\ell^-\gamma$ and $e^+e^-\to jj\gamma$ are studied, event selection strategies are discussed.
Based on the statistical significance, the expected constraints in experiments are estimated.
The required luminosities for the experiments to reach the unitarity bounds are presented.
\end{abstract}

%\begin{keywords}
%neutral triple gauge coupling, unitarity bound, CEPC
%\end{keywords}
%\keywords{neutral triple gauge coupling, unitarity bound, CEPC}

\maketitle
%\linenumbers

%\begin{multicols}{2}

\section{\label{level1}Introduction}

The extension of gauge interactions has been studied intensively in the searching of new physics~(NP) beyond the Standard Model~(SM).
In the SM effective field theory~(SMEFT) approach~\cite{weinberg,*SMEFTReview2,*SMEFTReview3,SMEFTReview1}, there are high-dimensional operators contributing to anomalous triple gauge couplings (aTGCs) and anomalous quartic gauge couplings~(aQGCs).
While the SMEFT has mainly been applied with only dimension-6 operators in the phenomenological studies, the importance of dimension-8 operators has been emphasised by many researchers~\cite{d81,*looportree,*ssww,*aqgcold,*aqgcnew,*Zhang:2018shp,*Bi:2019phv}.
It has been shown that the dimension-8 operators play a very important role from the convex geometry perspective to the SMEFT space~\cite{convexgeometry}.
Besides, there are processes sensitive to dimension-8 operators because the contributions from dimension-6 operators are absent~\cite{bi1,*bi2,*bi3}.
Some of the processes contributed by neutral triple gauge couplings~(nTGCs)~\cite{ntgc1,*ntgc4,*ntgc5,ntgcfcchh,ntgc2,ntgc3} are examples of such cases.

As an effective theory, the SMEFT is valid under a specific energy scale.
The contributions from high dimensional operators typically grow with the energy scale.
As a consequence, to probe the signals of high dimensional operators, one needs a large energy scale where the validity of the SMEFT becomes an important issue.
In the previous studies, the unitarity~\cite{unitarityHistory1,*unitarityHistory2,*unitarityHistory3,*unitaritynew1,*unitaritynew2} is often used to determine whether an effective theory is valid~\cite{unitarity1,*unitarity2,*unitarity3,*unitaritynew3,*unitaritynew4,*unitaritynew5,*unitaritynew6,*unitaritynew7}.
In a $e^+e^-$ collider such as the CEPC, without uncertainties from the parton distribution functions, the energy scale of a process $e^+e^-\to X$ is just the center mass~(c.m.) energy of the collider.
Therefore, the constraints on the coefficients of the nTGCs in the sense of unitarity can be obtained straightforwardly.
Meanwhile, the expected constraints at certain luminosity can be estimated by studying the signal significance.
For a fixed energy scale, there is a minimal luminosity that the constraints set by experiments are tighter than those required by unitarity.
Since the SMEFT is only valid within the unitarity bounds, the searching of the signals of nTGCs in experiments only makes sense when the luminosity is large enough.

In this work, we study the contribution of nTGCs in the process $e^+e^-\to \ell^+\ell^-\gamma$ with the focus on the case that $\ell=e,\mu$ and the $\ell^+\ell^-$ are from a $Z$ boson which has been studied in Refs.~\cite{ntgc2,ntgc3}, and the process $e^+e^-\to jj\gamma$ which has also been studied in Ref.~\cite{ntgc3}.
The partial wave unitarity bounds on nTGCs in this process at future $e^+e^-$ colliders are investigated.
The kinematic features and the event selection strategies for nTGCs at future $e^+e^-$ colliders are studied by using Monte-Carlo~(MC) simulation.
Different from Refs.~\cite{ntgc2,ntgc3}, the parton shower and fast detector simulation are applied, the event selection strategies are discussed.
With the help of signal significance, the required luminosities to study the nTGCs and the expected constraints at different colliders are investigated.

The rest of this paper is organized as follows, in Sec.~\ref{level2} we briefly introduce the operators contributing to nTGCs; the partial wave unitarity bounds are presented in Sec.~\ref{level3}; the numerical results based on MC simulation are shown in Sec.~\ref{level4}; finally Sec.~\ref{level5} is a summary.

\section{\label{level2}Dimension-8 operators contributing to nTGCs}

At dimension-8, there are $4$ CP-conserving operators contributing to nTGCs, they are~\cite{ntgc1,ntgc2}
\begin{equation}
\begin{split}
&\mathcal{L}_{\rm nTGC}=\frac{{\rm sign}(c_{\tilde{B}W})}{\Lambda_{\tilde{B}W}^4}\mathcal{O}_{\tilde{B}W}+\frac{{\rm sign}(c_{B\tilde{W}})}{\Lambda_{B\tilde{W}}^4}\mathcal{O}_{B\tilde{W}}
 +\frac{{\rm sign}(c_{\tilde{W}W})}{\Lambda_{\tilde{W}W}^4}\mathcal{O}_{\tilde{W}W}+\frac{{\rm sign}(c_{\tilde{B}B})}{\Lambda_{\tilde{B}B}^4}\mathcal{O}_{\tilde{B}B},\\
\end{split}
\label{eq.2.1}
\end{equation}
with
\begin{equation}
\begin{split}
 \mathcal{O}_{\tilde{B}W}=i H^{\dagger}\tilde{B}_{\mu \nu} W^{\mu \rho} \left\{D_{\rho },D^{\nu }\right\}H+h.c.,\;\;
&\mathcal{O}_{B\tilde{W}}=i H^{\dagger} B_{\mu \nu} \tilde{W}^{\mu \rho} \left\{D_{\rho },D^{\nu }\right\}H+h.c.,\\
 \mathcal{O}_{\tilde{W}W}=i H^{\dagger}\tilde{W}_{\mu \nu} W^{\mu \rho} \left\{D_{\rho },D^{\nu }\right\}H+h.c.,\;\;
&\mathcal{O}_{\tilde{B}B}=i H^{\dagger}\tilde{B}_{\mu \nu} B^{\mu \rho} \left\{D_{\rho },D^{\nu }\right\}H+h.c.,\\
\end{split}
\label{eq.2.2}
\end{equation}
where $H$ denotes the SM Higgs doublet, $\tilde{B}_{\mu \nu}\equiv \epsilon_{\mu\nu\alpha\beta} B^{\alpha\beta}$, $\tilde{W}_{\mu \nu}\equiv \epsilon_{\mu\nu\alpha\beta} W^{\alpha\beta}$ and $W_{\mu\nu}\equiv W_{\mu\nu}^a \sigma^{a}/2$ where $\sigma^a$ are Pauli matrices, $c_X$ are dimensionless coefficients, and the $\Lambda _X$ are related with the cutoff scale as $\Lambda _X= \Lambda / |c_X|^{1/4}$.
At leading order, the processes $e^+e^-\to \ell^+\ell^-\gamma$ and $e^+e^-\to jj\gamma$ can be affected by those operators via $ZV\gamma$ couplings where $V$ is a $Z$ boson or a photon.

It has been pointed out that, in the case of $ZV\gamma$ coupling with $Z$ boson and $\gamma$ on-shell, there is only one independent operator because the $\mathcal{O}_{B\tilde{W}}$ operator is equivalent to $\mathcal{O}_{\tilde{B}W}$ operator, $\mathcal{O}_{\tilde{W}W}$ and $\mathcal{O}_{\tilde{B}B}$ operators do not contribute~\cite{ntgc2}.
Therefore, in the following, we only consider the $\mathcal{O}_{\tilde{B}W}$ operator.

\section{\label{level3}The partial wave unitarity bound}

\begin{figure}[!htbp]
\centering{
\includegraphics[width=0.4\textwidth]{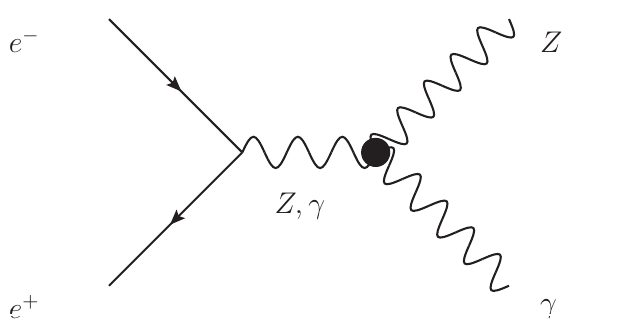}
\caption{\label{fig:fenydiag1}Feynman diagrams of the process $e^+e^-\to Z\gamma $ induced by nTGCs.}}
\end{figure}
When the contributions from nTGCs are taken into account, the cross-sections of the processes $e^+e^- \to \ell^+\ell^-\gamma$ and $e^+e^- \to jj\gamma$ grow with c.m. energy, which leads to the violation of unitarity at large enough energy.
The violation of unitarity indicates that the SMEFT is no longer valid to describe the phenomenon perturbatively, therefore partial wave unitarity is often used as a criterion to determine whether the SMEFT is valid.
In the case of $f\bar{f}\to V_1V_2$, where $f$ is a fermion, $\bar{f}$ is an anti-fermion, $V_{1,2}$ are vector bosons, the amplitude can be expanded as~\cite{partialwaveexpansion,*ffvv1}
\begin{equation}
\begin{split}
&\mathcal{M}(f_{\sigma _1}\bar{f}_{\sigma _2}\to V_{1,\lambda _3}V _{2,\lambda _4})
 =16\pi\sum _J \left(J+\frac{1}{2}\right)\delta _{\sigma_1,-\sigma _2}e^{i(m_1-m_2)\phi}d^J_{m_1,m_2}(\theta,\phi) T_J,\\
\end{split}
\label{eq.3.1}
\end{equation}
where $\sigma _{1,2}$ are helicities of the fermion and the anti-fermion, $\lambda _{1,2}$ are helicities of vector bosons, $m_1=\sigma _1-\sigma _2$, $m_2=\lambda _3-\lambda _4$, $d^J_{m_1,m_2}$ are Winger D functions, $\phi$ and $\theta$ are azimuth and zenith angles of $V_1$ and $T_J$ are coefficients of the partial wave expansion.
The partial wave unitarity bound is then $|T_J|\leq 1$~\cite{ffvv2}.

\begin{figure}[!htbp]
\centering{
\includegraphics[width=0.7\textwidth]{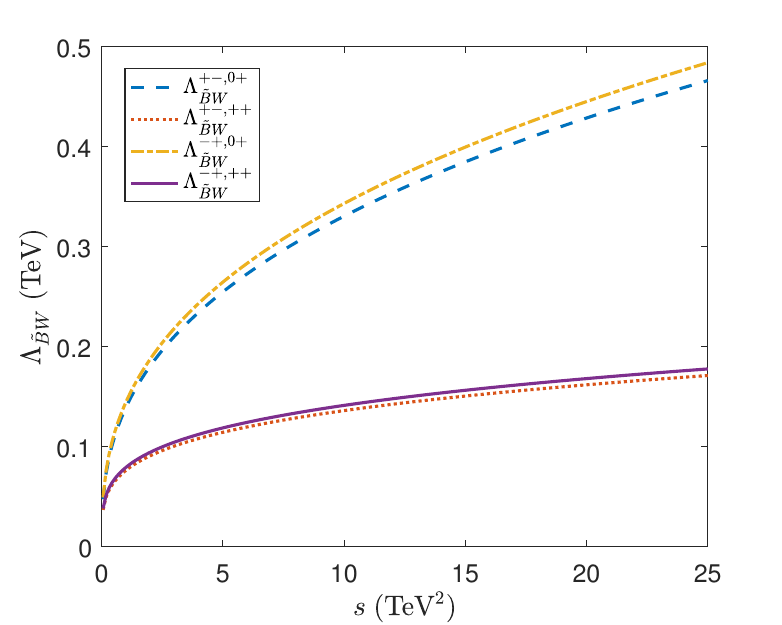}
\caption{\label{fig:unitaritybounds}The lower bounds of $\Lambda _{\tilde{B}W}$ from different helicity amplitudes.}}
\end{figure}

Since there is no nTGCs induced $e^+e^-\to \gamma\gamma$ with photons on-shell, we consider only $e^+e^-\to Z\gamma$. The Feynman diagrams of $e^+e^-\to Z\gamma$ induced by nTGCs are shown in Fig.~\ref{fig:fenydiag1}. The helical amplitudes can be obtained as
\begin{equation}
\begin{split}
&\mathcal{M}\left(e^-_{\frac{1}{2}}e^+_{-\frac{1}{2}}\to Z_0 \gamma _{\pm}\right)=\frac{e^2 e^{i \phi} \sqrt{s}\left[\frac{{\rm sign}\left(c_{\tilde{B}W}\right)v^2 \left( M_Z^2-s\right)^2 (\cos (\theta)\mp 1)}{\Lambda_{\tilde{B}W}^4}\pm 16M_z^2 s_W c_W\right]}{4 \sqrt{2} M_Z c_W^2\left(M_Z^2-s\right)},\\
&\mathcal{M}\left(e^-_{\frac{1}{2}}e^+_{-\frac{1}{2}}\to Z_{\pm} \gamma _{\pm}\right)= \pm \frac{e^2 e^{i \phi} \left[\frac{{\rm sign}\left(c_{\tilde{B}W}\right)}{\Lambda_{\tilde{B}W}^4}v^2 \left(M_Z^2-s\right)^2 \sin (\theta)-8M_z^2 c_Ws_W\left(\cot \left(\frac{\theta}{2}\right)\right)^{\pm 1}\right]}{4 c_W^2\left(M_Z^2-s\right)},\\
&\mathcal{M}\left(e^-_{\frac{1}{2}}e^+_{-\frac{1}{2}}\to Z_{\mp} \gamma _{\pm}\right)= \pm \frac{2e^{i\phi}e^2 s_W s \left(\tan \frac{\theta}{2}\right)^{\pm 1}}{c_W\left(s-M_z^2\right)},\\
&\mathcal{M}\left(e^-_{-\frac{1}{2}}e^+_{\frac{1}{2}}\to Z_0 \gamma _{\pm}\right)=\frac{1-2s_W^2}{2s_W^2}\frac{e^2 e^{-i \phi} \sqrt{s}\left[\frac{{\rm sign}\left(c_{\tilde{B}W}\right)v^2 \left( M_Z^2-s\right)^2 (\cos (\theta)\pm 1)}{\Lambda_{\tilde{B}W}^4}\pm 16M_z^2 s_W c_W\right]}{4 \sqrt{2} M_Z c_W^2\left(M_Z^2-s\right)},\\
&\mathcal{M}\left(e^-_{-\frac{1}{2}}e^+_{\frac{1}{2}}\to Z_{\pm} \gamma _{\pm}\right)=\pm \frac{1-2s_W^2}{2s_W^2}\frac{e^2 e^{-i \phi} \left[\frac{{\rm sign}\left(c_{\tilde{B}W}\right)}{\Lambda_{\tilde{B}W}^4}v^2 \left(M_Z^2-s\right)^2 \sin (\theta)+8M_z^2 c_Ws_W\left(\tan \left(\frac{\theta}{2}\right)\right)^{\pm 1}\right]}{4 c_W^2\left(M_Z^2-s\right)},\\
&\mathcal{M}\left(e^-_{-\frac{1}{2}}e^+_{\frac{1}{2}}\to Z_{\mp} \gamma _{\pm}\right)= \pm \frac{2s_W^2-1}{s_W^2}\frac{e^{-i\phi}e^2 s_W s \left(\cot \frac{\theta}{2}\right)^{\pm 1}}{c_W\left(s-M_z^2\right)}.\\
\end{split}
\label{eq.3.2}
\end{equation}
With Eqs.~(\ref{eq.3.1}), (\ref{eq.3.2}) and $|T_J|<1$, the unitarity bounds are
\begin{equation}
\begin{split}
&\left| \frac{{\rm sign}(c_{\tilde{B}W})}{ \Lambda_{\tilde{B}W}^{+-,0+} }-\frac{12 c_W s_W M_Z^2 }{v^2 \left(s-M_Z\right)^2} \right| < \frac{48 \sqrt{2} \pi  c_W^2 M_Z}{e^2 \sqrt{s} v^2 \left(s-M_Z^2\right)}\\
&\left| \frac{{\rm sign}(c_{\tilde{B}W})}{ \Lambda_{\tilde{B}W}^{+-,++} } - \frac{12 c_W s_W M_Z^2 }{v^2 \left(s-M_Z\right)^2}\right|<  \frac{48 \sqrt{2} \pi  c_W^2 }{e^2  v^2 \left(s-M_Z^2\right)}\\
&\left| \frac{{\rm sign}(c_{\tilde{B}W})}{ \Lambda_{\tilde{B}W}^{-+,0+} } + \frac{12 c_W s_W M_Z^2 }{v^2 \left(s-M_Z\right)^2} \right |<  \frac{96 \sqrt{2} \pi  s_W^2c_W^2M_Z}{e^2 v^2 \sqrt{s}(1-2s_W^2)\left(s-M_Z^2\right)}\\
&\left|  \frac{{\rm sign}(c_{\tilde{B}W})}{ \Lambda_{\tilde{B}W}^{-+,++} } + \frac{12 c_W s_W M_Z^2 }{v^2 \left(s-M_Z\right)^2} \right| <  \frac{96 \sqrt{2} \pi  c_W^2 s_W^2}{e^2  v^2 (1-2s_W^2)\left(s-M_Z^2\right)}\\
\end{split}
\label{eq.3.2add}
\end{equation}
For simplicity, we neglect the contributions from the SM which is small in the region of $\sqrt{s}$ we are interested in, then
\begin{equation}
\begin{split}
 \Lambda_{\tilde{B}W}^{+-,0+}\geq \left(\frac{e^2  \sqrt{s} v^2 \left(s-M_Z^2\right)}{48 \sqrt{2} \pi  M_Z c_W^2}\right)^{\frac{1}{4}},\;\;
&\Lambda_{\tilde{B}W}^{+-,++}\geq \left(\frac{e^2  v^2 \left(s-M_Z^2\right)}{48 \sqrt{2} \pi  c_W^2}\right)^{\frac{1}{4}},\\
 \Lambda_{\tilde{B}W}^{-+,0+}\geq \left(\frac{e^2  \sqrt{s} \left(1-2 s_W^2\right) v^2 \left(s-M_Z^2\right)}{96 \sqrt{2} \pi  M_Z s_W^2 c_W^2}\right)^{\frac{1}{4}},\;\;
&\Lambda_{\tilde{B}W}^{-+,++}\geq \left(\frac{e^2  \left(2 s_W^2-1\right) v^2 \left(M_Z^2-s\right)}{96 \sqrt{2} \pi  s_W^2 c_W^2}\right)^{\frac{1}{4}},\\
\end{split}
\label{eq.3.3}
\end{equation}
and bounds from $e^-_{\frac{1}{2}}e^+_{-\frac{1}{2}} \to Z_0\gamma _-$, $e^-_{\frac{1}{2}}e^+_{-\frac{1}{2}} \to Z_-\gamma _-$,$e^-_{-\frac{1}{2}}e^+_{\frac{1}{2}} \to Z_0\gamma _-$ and $e^-_{-\frac{1}{2}}e^+_{\frac{1}{2}} \to Z_-\gamma _-$ are same as those from $e^-_{\frac{1}{2}}e^+_{-\frac{1}{2}} \to Z_0\gamma _+$, $e^-_{\frac{1}{2}}e^+_{-\frac{1}{2}} \to Z_+\gamma _+$,$e^-_{-\frac{1}{2}}e^+_{\frac{1}{2}} \to Z_0\gamma _+$ and $e^-_{-\frac{1}{2}}e^+_{\frac{1}{2}} \to Z_+\gamma _+$, respectively. The unitarity bounds are depicted in Fig.~\ref{fig:unitaritybounds}.

Since there is always $s>M_Z^2$, the strongest bound is ${\Lambda ^{-+,0+}_{\tilde{B}W}}$.
We consider $\sqrt{s}=250, 500, 1000, 3000\;{\rm GeV}$ which approximately correspond to the expected c.m. energies of the FCC-ee~\cite{FCCee} and CEPC~\cite{CEPC1,*CEPC2}, ILC~\cite{ILC}, ILC upgrades and the CLIC~\cite{CLIC}, respectively, and the case of $\sqrt{s}=5000\;{\rm GeV}$ is also included which has been investigated for nTGCs~\cite{ntgc2}.
The unitarity bounds are listed in Table.~\ref{Tab:unitaritybound}
\begin{table*}
\begin{center}
\begin{tabular}{c|c|c|c|c|c}
\hline
$\sqrt{s}$ & $250$ GeV & $500$ GeV & $1000$ GeV & $3000$ GeV & $5000$ GeV \\
\hline
$\Lambda_{\tilde{B}W}$ & $>49.41$ GeV & $>85.38$ GeV & $>144.52$ GeV & $>330.04$ GeV & $>484.19$ GeV \\
\hline
\end{tabular}
\end{center}
\caption{\label{Tab:unitaritybound}The constraints on $\Lambda _{\tilde{B}W}$ from unitarity bounds.}
\end{table*}

\section{\label{level4}Numerical study}

The coefficients of nTGCs can be constrained by experiments.
However, only when the constraints are tighter than the unitarity bounds, the SMEFT is valid and the constraints make sense.
By analysing the signal significance, one can estimate the minimal luminosities for the experiments so that the constraints obtained can be tighter than the unitarity bounds.
In this section, the numerical results obtained by MC simulation are presented.

\subsection{\label{level4.1}Leptonic Z decays}

In the following, we combine the processes $e^+e^-\to e^+e^-\gamma$ and $e^+e^-\to \mu^+\mu^-\gamma$ as $e^+e^-\to \ell^+\ell^-\gamma$. This process can also been affected by dimension-6 operators. We choose the Warsaw basis~\cite{SMEFTReview1,warsawbase2} to discuss the contribution from the dimension-6 operators to this process.
For most Feynman diagrams, the $\ell^{\pm}$ are not from a $Z$ resonance, therefore to concentrate on the signals of nTGCs, one can apply a cut on the invariant mass of $\ell^{\pm}$.
Except for that, the contributions from the dimension-8 operators grow faster with the energy than the dimension-6 operators.
Besides, when the $\ell^{\pm}$ are from a $Z$ resonance, the angular distributions can also be used to discriminate the signals of nTGCs from the signals of dimension-6 operators.
T-channel $e^+e^-\to Z\gamma$ diagrams can be induced by the $\ell^+\ell^- V$ vertices provided by operators such as $\mathcal{O}_{H \ell}^{(1)}=(H^{\dagger} i\overleftrightarrow {D} _{\mu}H) (\bar{L} '_p)\gamma ^{\mu} L ' _r$.
Here we use $L$ to denote the leptons including $\tau$ which is different from $\ell$.
However, the diagrams induced by nTGCs are s-channel diagrams, therefore the angular distributions of $Z$ boson and photon should be different.
Meanwhile, an s-channel $e^+e^- \to H \to Z\gamma$ diagram can be induced by the $HZ\gamma$ vertex provided by $\mathcal{O}_{H WB}=H^{\dagger} \sigma^a H W_{\mu\nu}^a B_{\mu\nu}$.
The production of Higgs boson should be small since the coupling is proportional to the lepton mass, and this background can be easily removed by cutting off the events with the invariant mass of $\ell^{\pm},\gamma$ near the mass of Higgs boson.
Besides, since the Higgs boson is a scalar, the polarization feature of the $Z$ boson in the final state is different from the case of nTGCs, where the $Z$ bosons are dominantly longitudinal polarized as will be shown later.
As a result, the angular distributions of $\ell^{\pm}$ are different.
Nevertheless, since we concentrate on the sensitivity of this process to the nTGCs, we assume one operator at a time and neglect the contributions from dimension-6 operators.

The features of the signals and backgrounds are studied by using the \verb"MadGraph5_aMC@NLO" toolkit~\cite{madgraph,*feynrules}.
The fast detector simulation is applied by using \verb"Delphes"~\cite{delphes} with the CEPC detector card.
The basic cuts are set as same as the default settings except for $\Delta R_{\ell\ell}$ which is defined as $\sqrt{\Delta \eta_{\ell\ell} ^2+\Delta \phi_{\ell\ell}^2}$ where $\Delta \eta_{\ell\ell}$ and $\Delta \phi_{\ell\ell}$ are the differences between pseudo-rapidities and azimuth angles of the leptons, respectively.
The kinematic features are studied using \verb"MLAnalysis"~\cite{Guo:2023nfu}. 
As will be explained later, in the basic cuts we use $\Delta R_{\ell\ell}>0.2$.
In the MC simulation, we use the coefficients in the ranges listed in Table~\ref{Tab:region}.

\begin{figure}[!htbp]
\centering{
\includegraphics[width=0.7\textwidth]{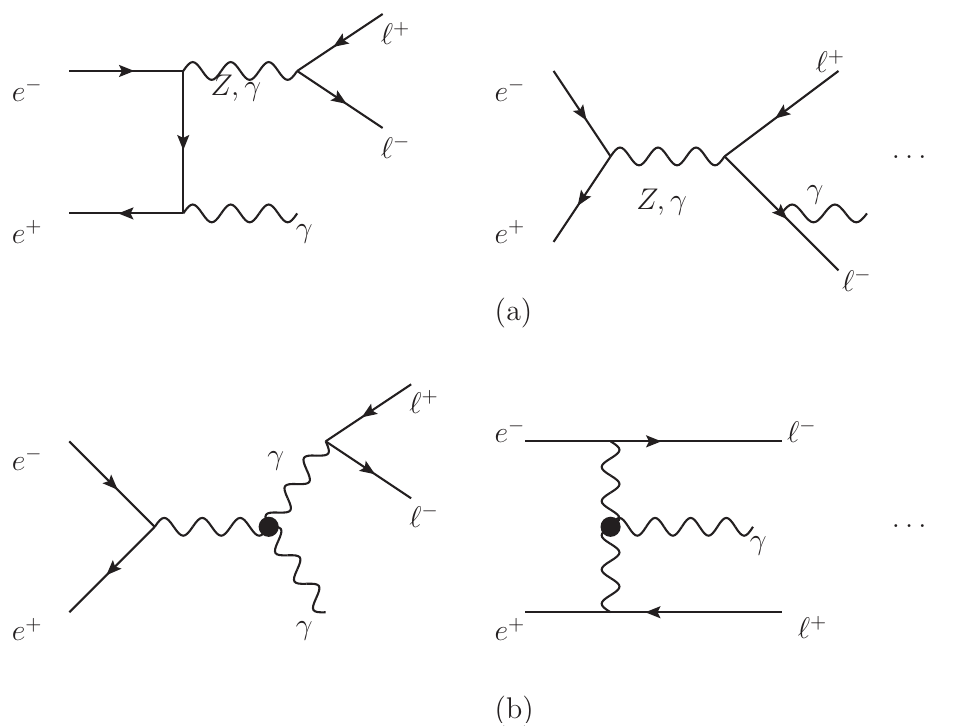}
\caption{\label{fig:fenydiag2}Feynman diagrams which contribute to the process $e^+e^-\to \ell ^+\ell ^-\gamma $}}
\end{figure}

We consider the process $e^+e^-\to \ell^+\ell^-\gamma $, the dominant signal of the $\mathcal{O}_{\tilde{B}W}$ operator is from the diagrams depicted in Fig.~\ref{fig:fenydiag1} joint with $Z\to \ell^+\ell^-$. The SM backgrounds are depicted in Fig.~\ref{fig:fenydiag2}.~(a).
To study the kinematic features of the dominant signal, the signal events are generated with the largest coefficients in Table~\ref{Tab:region}.

\begin{table*}
\begin{center}
\begin{tabular}{c|c|c|c|c}
 $\sqrt{s}=250$ GeV & $\sqrt{s}=500$ GeV & $\sqrt{s}=1$ TeV & $\sqrt{s}=3$ TeV & $\sqrt{s}=5$ TeV\\
\hline
 $[-1000,1000]$ & $[-200,200]$ & $[-30,30]$ & $[-2,2]$ & $[-0.8,0.8]$ \\
\end{tabular}
\end{center}
\caption{\label{Tab:region}The ranges of coefficients ${\rm sign(c_{\tilde{B}W})}/\Lambda _{\tilde{B}W}^4$ (${\rm TeV}^{-4}$) used in the MC simulation.}
\end{table*}

We require the particle numbers in the final states to be $N_{\ell^+}\geq 1$, $N_{\ell^-}\geq 1$ and $N_{\gamma}\geq 1$.
Moreover, we require the hardest two leptons to be a lepton and an anti-lepton with the same flavor.
This requirement is denoted as $N_{\ell,\gamma}$ cut, in the following, results are presented after $N_{\ell,\gamma}$ cut.

\begin{figure*}[!htbp]
\centering{
\subfigure[]{\includegraphics[width=0.49\textwidth]{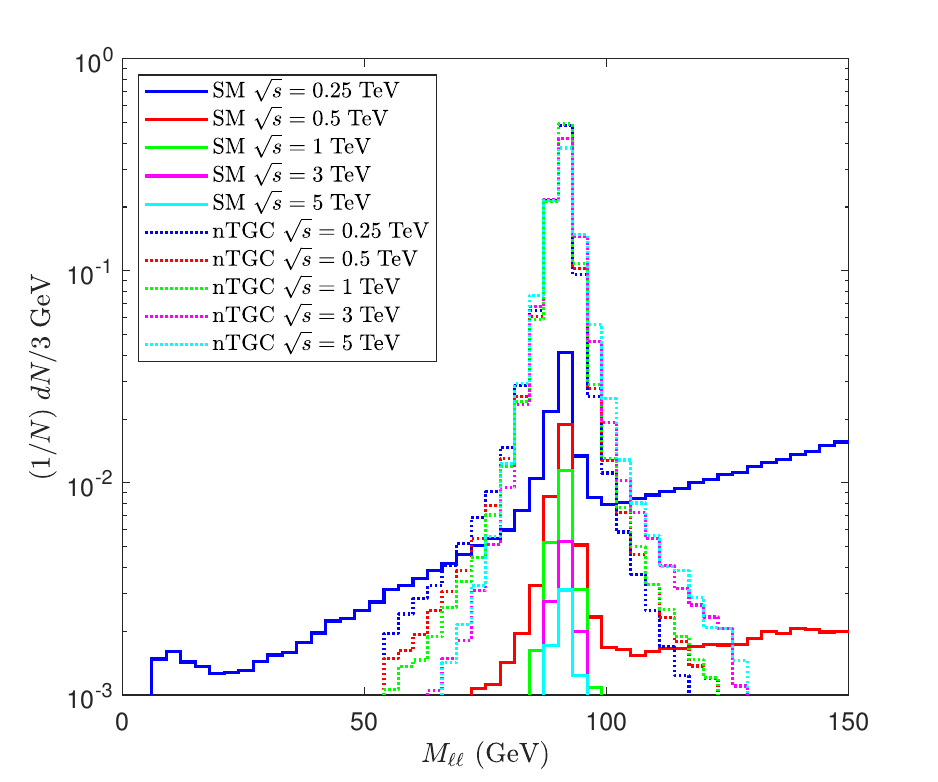}}
\subfigure[]{\includegraphics[width=0.49\textwidth]{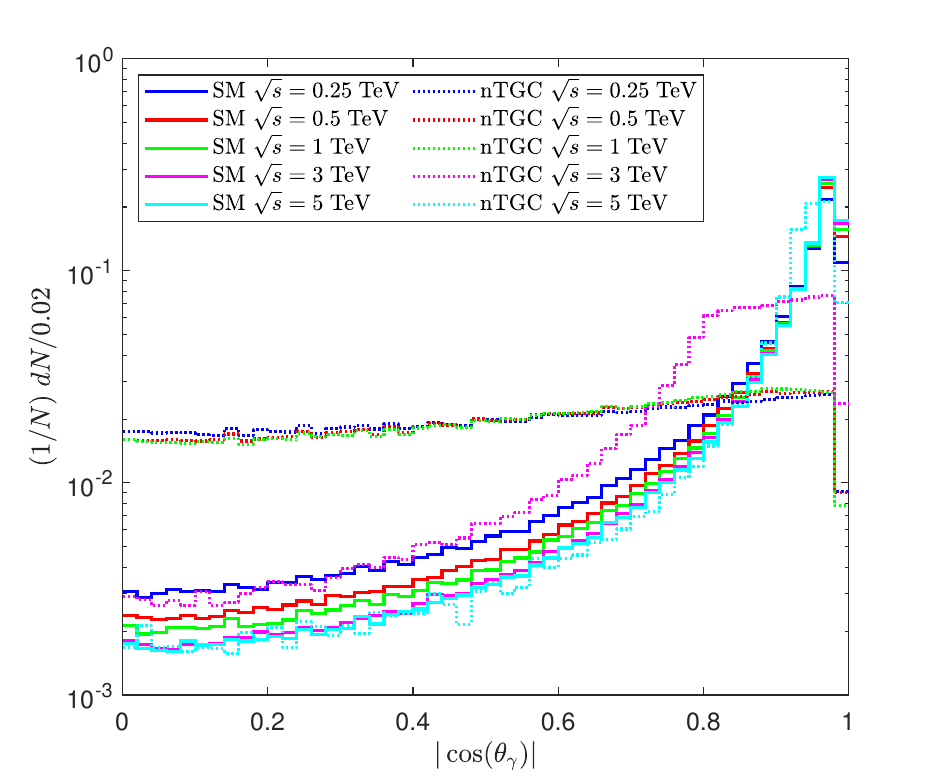}}\\
\subfigure[]{\includegraphics[width=0.49\textwidth]{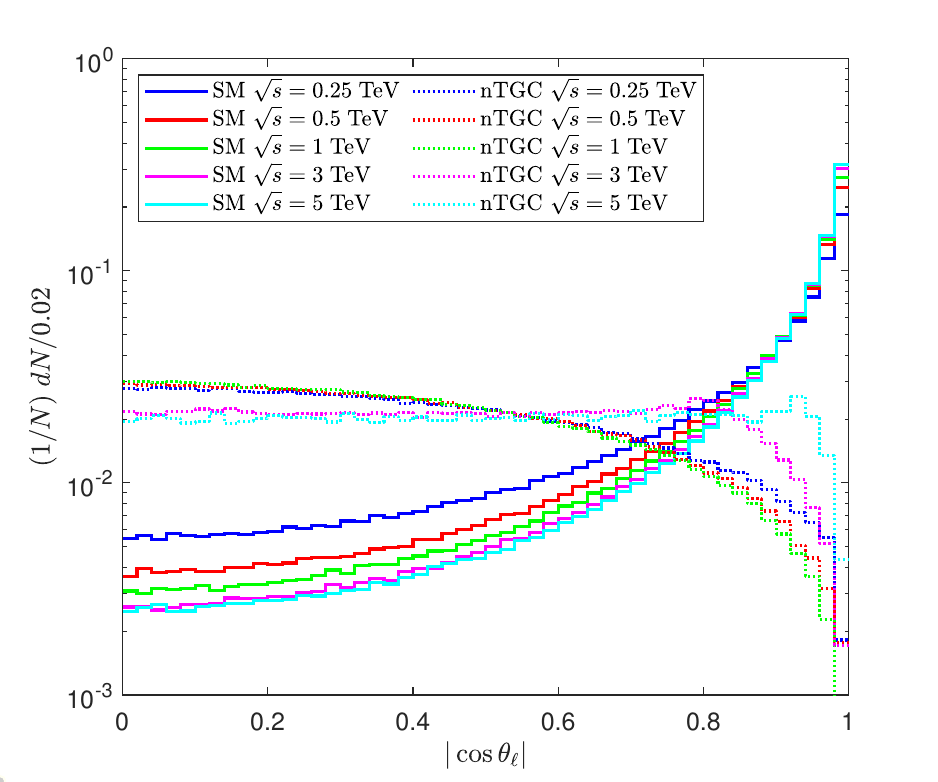}}
\subfigure[]{\includegraphics[width=0.49\textwidth]{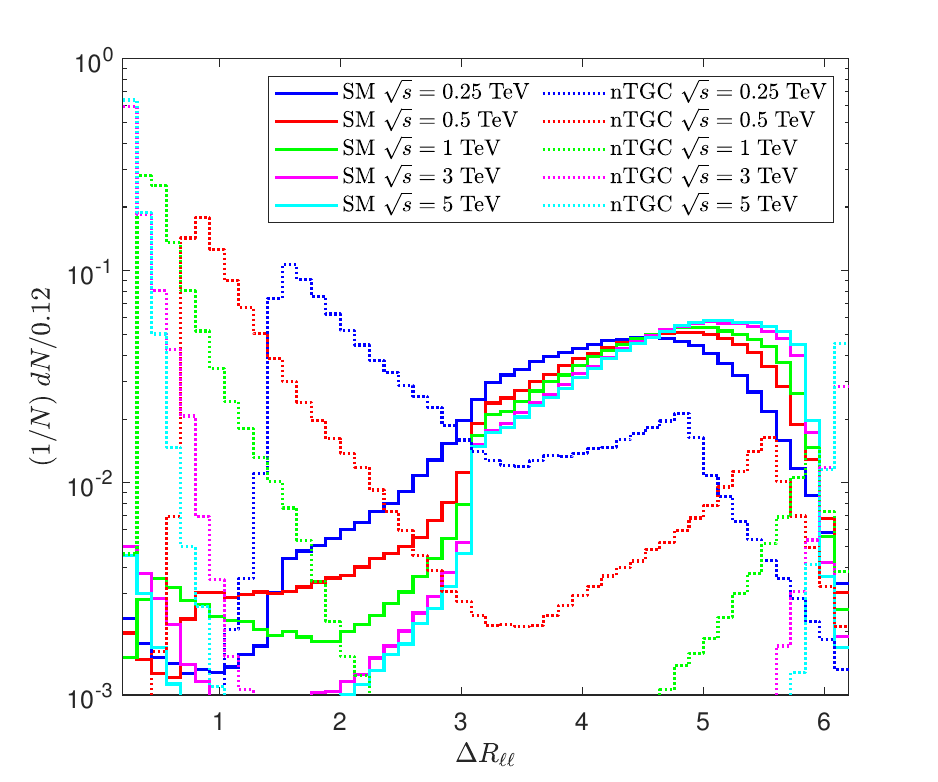}}
\caption{\label{Fig:KinematicFeature}The normalized distributions of $M_{\ell\ell}$, $|\cos \theta _{\gamma}|$, $|\cos \theta _{\ell}|$ and $\Delta R_{\ell\ell}$.}}
\end{figure*}

To remove the backgrounds without a $Z$ resonance, we require the invariant mass of the leptons~(denoted as $M_{\ell\ell}$) to be close to $M_Z$.
The normalized distributions of $M_{\ell\ell}$ are shown in Fig.~\ref{Fig:KinematicFeature}.~(a).
In the distributions of the signal, the peaks at $M_Z$ are much sharper than those in the distributions of the SM backgrounds.
Defining $\Delta M=|M_{\ell\ell} - M_Z|$, we cut off the events with $\Delta M>15$ GeV.

In the signal events, the photons are emitted from the s-channel diagrams.
Therefore, in the c.m. frame of $e^+e^-$, the zenith angles of the photons~(denoted as $\theta _{\gamma}$) are different from those emitted from the $e^{\pm}$ in the initial state, because typically the contribution from a t-channel diagram inversely proportional to the square of Mandelstam variable $t$ which goes to $0$ when $\theta _{\gamma}\to 0$.
$\theta _{\gamma}$ has been proposed to discriminate the signal from the backgrounds also in the previous study~\cite{ntgc2}.
The normalized distributions of $|\cos (\theta _{\gamma})|$ are shown in Fig.~\ref{Fig:KinematicFeature}.~(b).
We cut off the events with a large $|\cos (\theta _{\gamma})|$.

The polarizations of $e^{\pm}$ beams have been studied in Refs.~\cite{ntgc2,ntgc3}.
Since the longitudinal polarization is difficult to realize at the circular colliders, instead we consider the polarization effect in the final state which has been used to highlight the signals in other scenarios~\cite{wastudy,ntgc2,zzpolarization,*gggpolarization,*dijetgpolarization}.
One can see from Eq.~(\ref{eq.3.2}) that the $Z$ bosons in the final states are dominantly longitudinal polarized at large $\sqrt{s}$.
This leads to a unique angular distribution of the leptons in the rest frame~(the helical frame) of the $Z$ boson.
In the rest frame of $\ell^+\ell^-$ with ${\bf z}$-axis pointing to the direction of ${\bf p}_{\ell^+}+{\bf p}_{\ell^-}$, and with the zenith angle of the lepton in this frame denoted as $\theta _{\ell}$, the normalized distributions of $|\cos (\theta _{\ell})|$ are shown in Fig.~\ref{Fig:KinematicFeature}.~(c).
We cut off the events with a large $|\cos (\theta _{\ell})|$.

Another important issue is $\Delta R_{\ell\ell}$. When the $Z$ boson is energetic, the leptons are approximately collinear to each other.
As a result, the $\Delta R_{\ell\ell}$ is small for the signal events.
However, $\Delta R_{\ell\ell}$ is related to the isolation of leptons, therefore a none zero $\Delta R_{\ell\ell}$ is required in the experiments.
To keep the signal events, the lower bound of $\Delta R_{\ell\ell}$ should be as small as possible~\cite{zgamma}.
Therefore, in the basic cuts we use $\Delta R_{\ell\ell}>0.2$ which can be realised in experiments~\cite{zaexp1}.
Note that the cross-section and the kinematic features are greatly affected by this cut.
For example, at $\sqrt{s}=3$ TeV and $\sqrt{s}=5$ TeV, with $\Delta R_{\ell\ell}>0.2$ in basic cuts and with $N_{\ell,\gamma}$ cut, approximately $72\%$ and $90\%$ signal events are lost, respectively, compared with $\sigma(\ell^+\ell^-\gamma)=\sigma(Z\gamma)\times {\rm Br}(Z\to \ell^+\ell^-)$.
The normalized distributions of $\Delta R_{\ell\ell}$ are shown in Fig.~\ref{Fig:KinematicFeature}.~(d).
We find that, for the SM backgrounds, typically $\Delta R_{\ell\ell}$ lies in the region $2<\Delta R_{\ell\ell}<6$.
We keep the events with $\Delta R_{\ell\ell}$ outside of this region.

\begin{table*}
\begin{center}
\begin{tabular}{c|c|c|c|c|c|c|c|c|c|c}
 & \multicolumn{2}{c|}{250 GeV} & \multicolumn{2}{c|}{500 GeV} & \multicolumn{2}{c|}{1 TeV} & \multicolumn{2}{c|}{3 TeV} & \multicolumn{2}{c}{5 TeV} \\
\hline
 & SM & nTGC & SM & nTGC & SM & nTGC & SM & nTGC & SM & nTGC \\
 \hline
$N_{\ell,\gamma}$ cut & $5490.5$ & $390.8$     & $2158.4$ & $303.6$     & $754.6$ & $112.9$      & $123.7$ & $15.8$ & $50.6$ & $6.91$ \\
\hline
$\Delta M<15$ GeV & $721.7$ & $373.3$      & $99.7$ & $291.6$     & $19.1$ & $109.0$      & $1.60$ & $15.2$ & $0.41$ & $6.65$ \\
\hline
$|\cos(\theta _{\gamma})|<0.9$ & $306.0$ & $331.8$      & $48.0$ & $257.9$     & $10.2$ & $96.3$      & & & & \\
$|\cos(\theta _{\gamma})|<0.95$ & & & & & & & $1.07$ & $13.1$ & & \\
\hline
$|\cos(\theta _{\ell})|<0.8$ & $217.5$ & $313.6$     & $35.8$ & $240.7$     & $7.77$ & $90.5$      & & & & \\
$|\cos(\theta _{\ell})|<0.9$ & & & & & & & $0.90$ & $12.6$ & & \\
$|\cos(\theta _{\ell})|<0.95$ & & & & & & & & & $0.37$ & $6.47$ \\
\hline
$\Delta R_{\ell\ell}\notin [0.7, 6]$ &  &  &  &  &  &  & $0.87$ & $12.4$ & & \\
$\Delta R_{\ell\ell}\notin [0.5, 6]$ &  &  &  &  &  &  & & & $0.31$ & $6.21$ \\
\end{tabular}
\end{center}
\caption{\label{Tab:cutflow}The event selection strategy and cross-sections~(fb) after cuts.}
\end{table*}

The proposed event selection strategy and the cross-sections after cuts are summarised in Table~\ref{Tab:cutflow}.
The SM backgrounds can be effectively reduced by our selection strategy.

\begin{figure*}[!htbp]
\centering{
\subfigure[$\sqrt{s}=250$ GeV]{\includegraphics[width=0.32\textwidth]{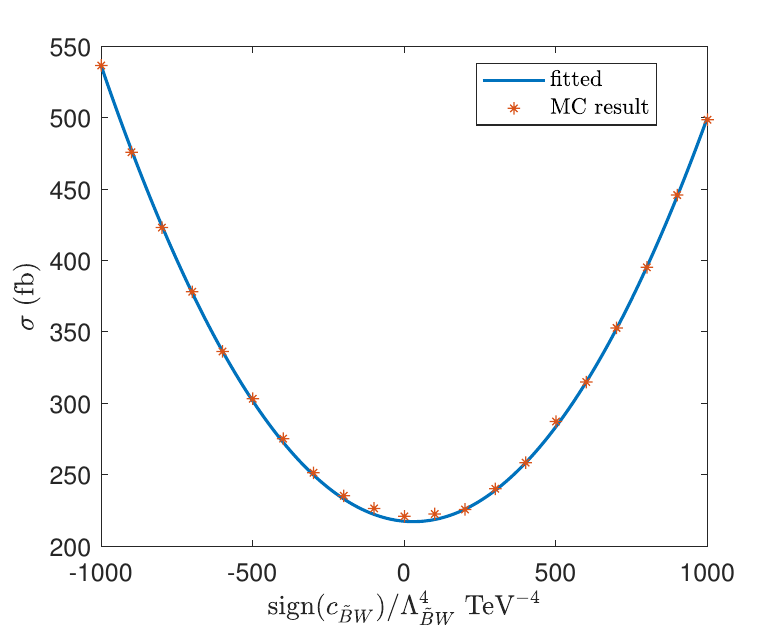}}
\subfigure[$\sqrt{s}=500$ GeV]{\includegraphics[width=0.32\textwidth]{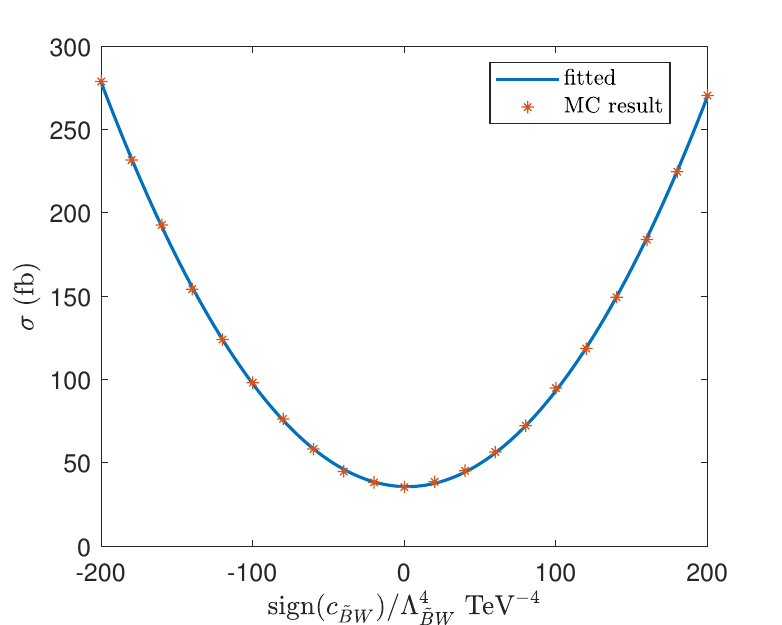}}
\subfigure[$\sqrt{s}=1$ TeV]{\includegraphics[width=0.32\textwidth]{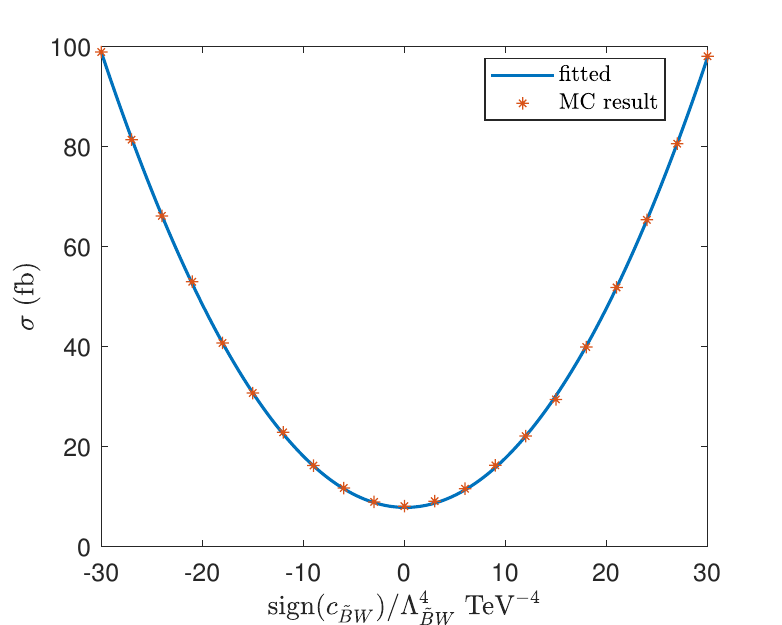}}\\
\subfigure[$\sqrt{s}=3$ TeV]{\includegraphics[width=0.32\textwidth]{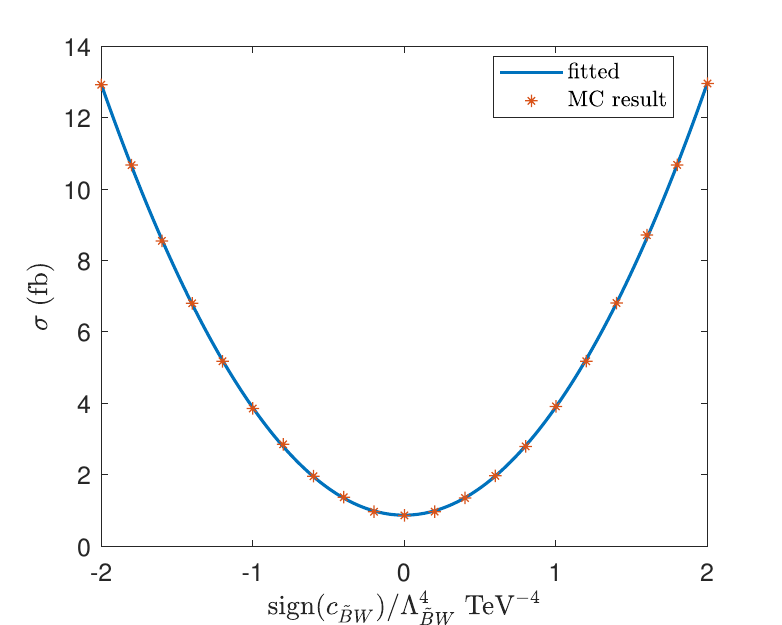}}
\subfigure[$\sqrt{s}=5$ TeV]{\includegraphics[width=0.32\textwidth]{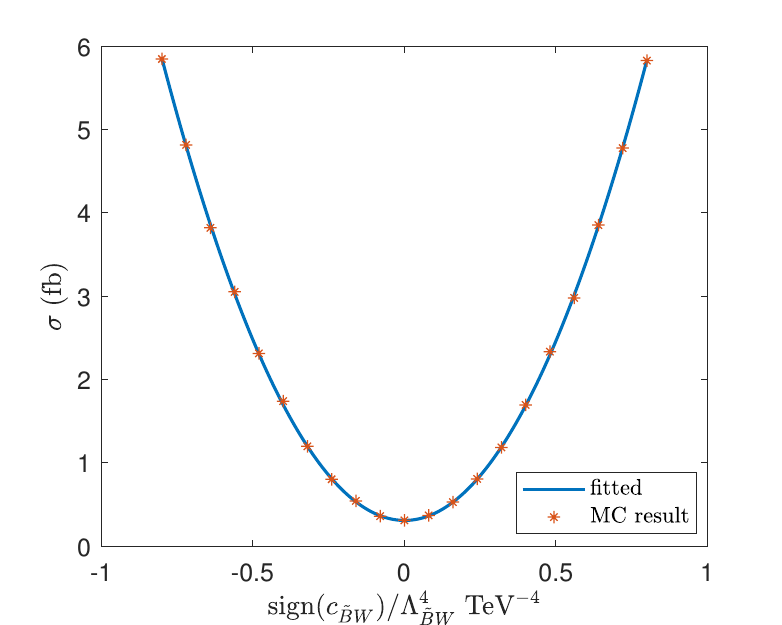}}
\caption{\label{Fig:cs}Cross-sections as functions of ${\rm sign}(c_{\tilde{B}W})/\Lambda ^4_{\tilde{B}W}$.}}
\end{figure*}

\begin{figure*}[!htbp]
\centering{
\subfigure[$\sqrt{s}=250$ GeV]{\includegraphics[width=0.32\textwidth]{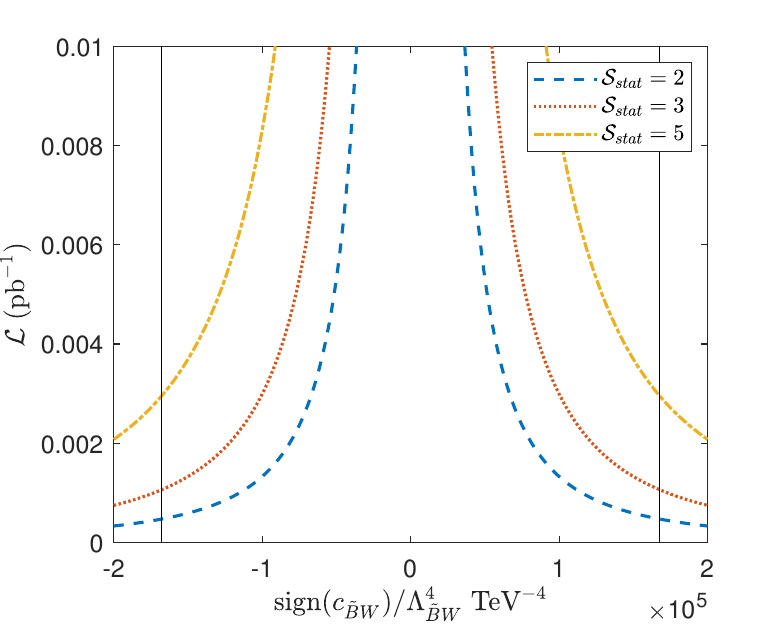}}
\subfigure[$\sqrt{s}=500$ GeV]{\includegraphics[width=0.32\textwidth]{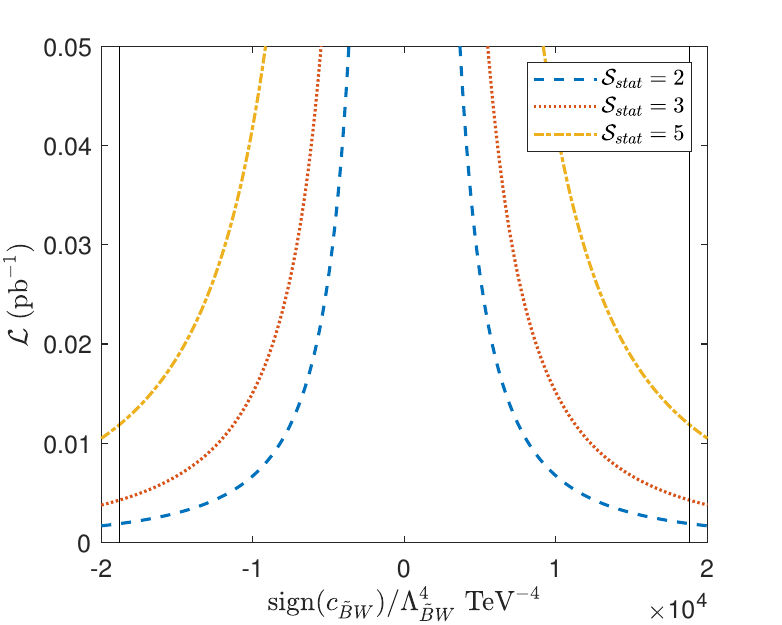}}
\subfigure[$\sqrt{s}=1$ TeV]{\includegraphics[width=0.32\textwidth]{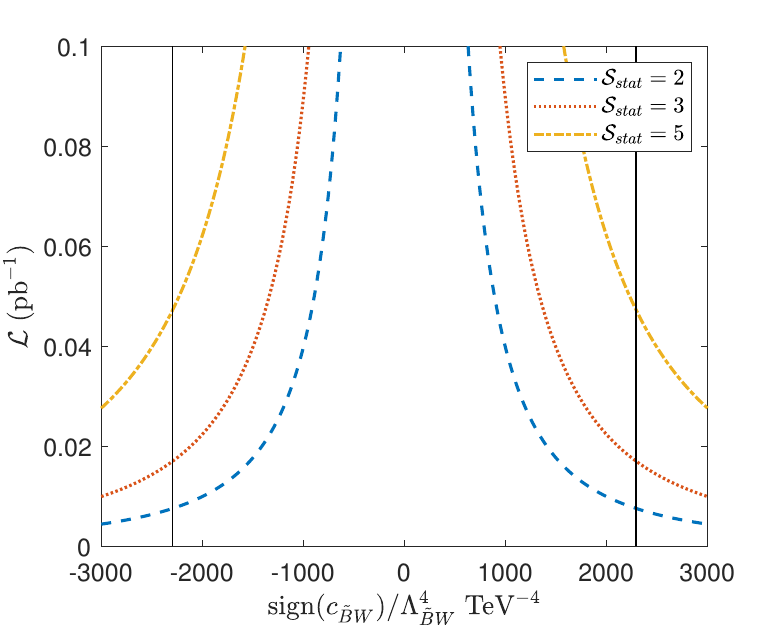}}\\
\subfigure[$\sqrt{s}=3$ TeV]{\includegraphics[width=0.32\textwidth]{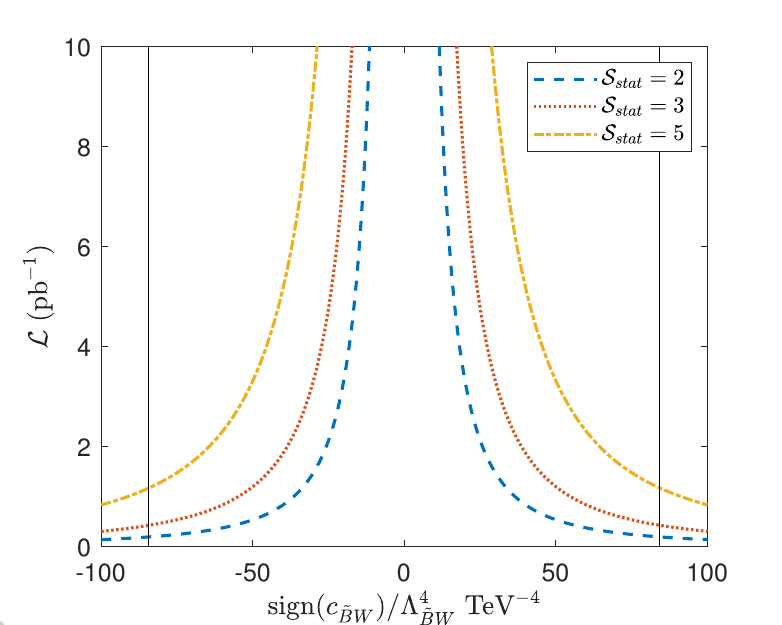}}
\subfigure[$\sqrt{s}=5$ TeV]{\includegraphics[width=0.32\textwidth]{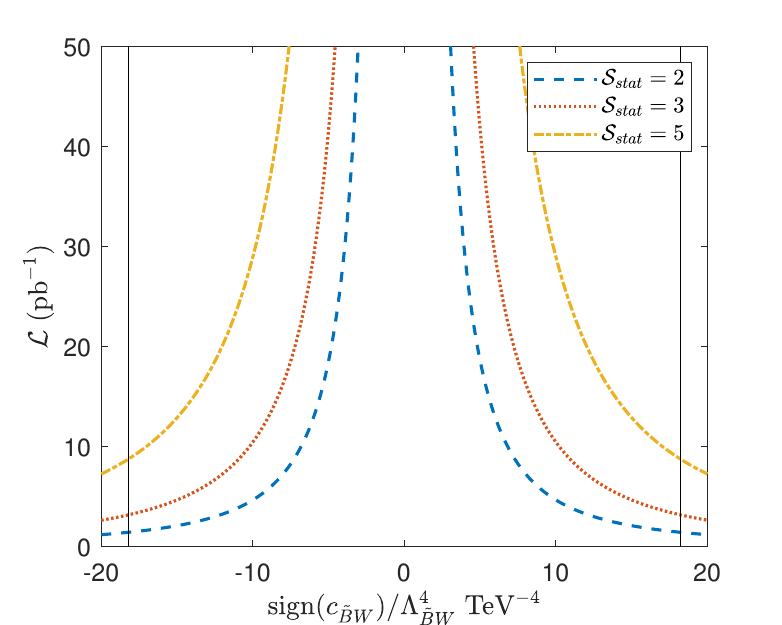}}
\caption{\label{Fig:ss}The constraints corresponding to different luminosities. The vertical lines are unitarity bounds.}}
\end{figure*}

\begin{table*}
\begin{center}
\begin{tabular}{c|c|c|c|c}
 $\sqrt{s}=250$ GeV & $\sqrt{s}=500$ GeV & $\sqrt{s}=1$ TeV & $\sqrt{s}=3$ TeV & $\sqrt{s}=5$ TeV\\
\hline
 $0.0030\;{\rm pb}^{-1}$ & $0.012\;{\rm pb}^{-1}$ & $0.047\;{\rm pb}^{-1}$ & $1.2\;{\rm pb}^{-1}$ & $8.7\;{\rm pb}^{-1}$ \\
\end{tabular}
\end{center}
\caption{\label{Tab:lumolosity}The required luminosities for the constraints obtained by experiments at $\mathcal{S}_{stat}=5$ to be tighter than the unitarity bounds.}
\end{table*}

By scanning the parameter spaces listed in Table~\ref{Tab:region}, the cross-sections of the process $e^+e^-\to \ell^+\ell^- \gamma$ at different coefficients are obtained.
To study how the process is affected by nTGCs, the diagrams in Fig.~\ref{fig:fenydiag2}.~(b) and interference terms are included.
The cross-sections are bilinear functions of coefficients.
As shown in Fig.~\ref{Fig:cs} that the numerical results fit the bilinear functions well.
The symmetry axes of the parabolas are close to zero, indicating that the interference between NP and the SM is negligible within the ranges of coefficients we use.
The constraints on the coefficient can be estimated by using signal significance defined as $\mathcal{S}_{stat}=N_{\rm nTGC}/\sqrt{N_{\rm nTGC}+N_{\rm sm}}$ where $N_{\rm nTGC}$ is the number of signal events, and $N_{\rm sm}$ is the number of events of the SM backgrounds.
The constraints corresponding to different luminosities~(denoted as $\mathcal{L}$) are shown in Fig.~\ref{Fig:ss}.
As a compare, the unitarity bounds are also shown in Fig.~\ref{Fig:ss} as vertical lines.
Considering the constraints set by experiments at $\mathcal{S}_{stat}=5$, the required luminosities for the constraints to be tighter than the unitarity bounds are listed in Table~\ref{Tab:lumolosity}.
Using $\mathcal{L}=2\;{\rm ab}^{-1}$ as a representation, the expected constraints in experiments are shown in Table~\ref{Tab:constraints}.
Note that, in Table~\ref{Tab:constraints}, the ranges of expected restrictions for coefficients are reduced by two orders of magnitude than Table~\ref{Tab:region}.
For the range in Table~\ref{Tab:constraints} and for the cases of $\sqrt{s}=250\;{\rm GeV}$ and $500\;{\rm GeV}$, the asymmetries of the expected restrictions indicate the importance of the interference terms.
To study the phenomenon around these ranges of coefficients, the $\phi ^*$ can be used to highlight the contribution from interference terms which is defined as the angle between the scattering plane and the decay plane of Z in the rest frame of Z~\cite{ntgc2,ntgc3} and the signal significant can be further improved.
Compare our result with the studies of other dimension-8 operators at the LHC~\cite{wastudy,zaexp3,*coefficient1,*Guo:2019agy}, and the study of nTGCs at FCC-hh~\cite{ntgcfcchh}, the process $e^+e^-\to\ell^+\ell^-\gamma$ at the $e^+e^-$ colliders shows competitive sensitivity to dimension-8 operators.

\begin{table*}
\begin{center}
\begin{tabular}{c|c|c|c|c|c}
 $\mathcal{S}_{stat}$ & $\sqrt{s}=250$ GeV & $\sqrt{s}=500$ GeV & $\sqrt{s}=1$ TeV & $\sqrt{s}=3$ TeV & $\sqrt{s}=5$ TeV\\
\hline
 $2$ & $[-25.7, 85.4]$ & $[-5.2, 8.7]$ & $[-1.0, 1.2]$ & $[-0.12, 0.12]$ & $[-0.054, 0.056]$ \\
 $3$ & $[-34.9, 94.6]$ & $[-6.7, 10.2]$ & $[-1.3, 1.5]$ & $[-0.15, 0.15]$ & $[-0.067, 0.069]$ \\
 $5$ & $[-50.1, 109.8]$ & $[-9.0, 12.5]$ & $[-1.7, 1.9]$ & $[-0.19, 0.19]$ & $[-0.088, 0.090]$ \\
\end{tabular}
\end{center}
\caption{\label{Tab:constraints}The constraints on ${\rm sign(c_{\tilde{B}W})}/\Lambda _{\tilde{B}W}^4$ (${\rm TeV}^{-4}$) at $\mathcal{L}=2\;{\rm ab}^{-1}$.}
\end{table*}

\subsection{\label{level4.2}Hadronic Z decays}

\begin{figure}[!htbp]
\centering{
\includegraphics[width=0.7\textwidth]{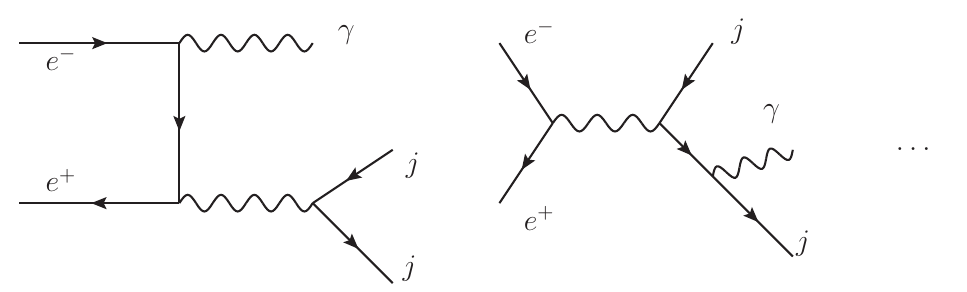}
\caption{\label{fig:fenydiag3}Feynman diagrams which contribute to the process $e^+e^-\to jj\gamma $}}
\end{figure}

To study the hadronic Z decays, the events are showed by using \verb"Pythia8"~\cite{pythia8} before the fast detector simulation.
We only consider the existence of the $\mathcal{O}_{\tilde{B}W}$ operator.
Typical diagrams of the SM backgrounds are depicted in Fig.~\ref{fig:fenydiag3}.

Similar as the problem of $\Delta R_{\ell\ell}$, when the $Z$ boson is energetic, the jets tend to be co-linear to each other, which has an important effect on the result~\cite{ntgc3}.
For example, for the case of $\sqrt{s}=5\;({\rm TeV})$, after a fast detector simulation using anti-${\rm k}_T$ algorithm with a cone radius $R=0.5$, and $p_{T,min}=20\;{\rm GeV}$, about $99.6\%$ events are lost if one requires two jets in the final state.
Therefore, we require the particle numbers in the final states to be $N_j\geq 1, N_{\gamma}\geq 1$ which is denoted as $N_{j,\gamma}$ cut, in the following, results are presented after $N_{j,\gamma}$ cut.

\begin{figure*}[!htbp]
\centering{
\subfigure[]{\includegraphics[width=0.49\textwidth]{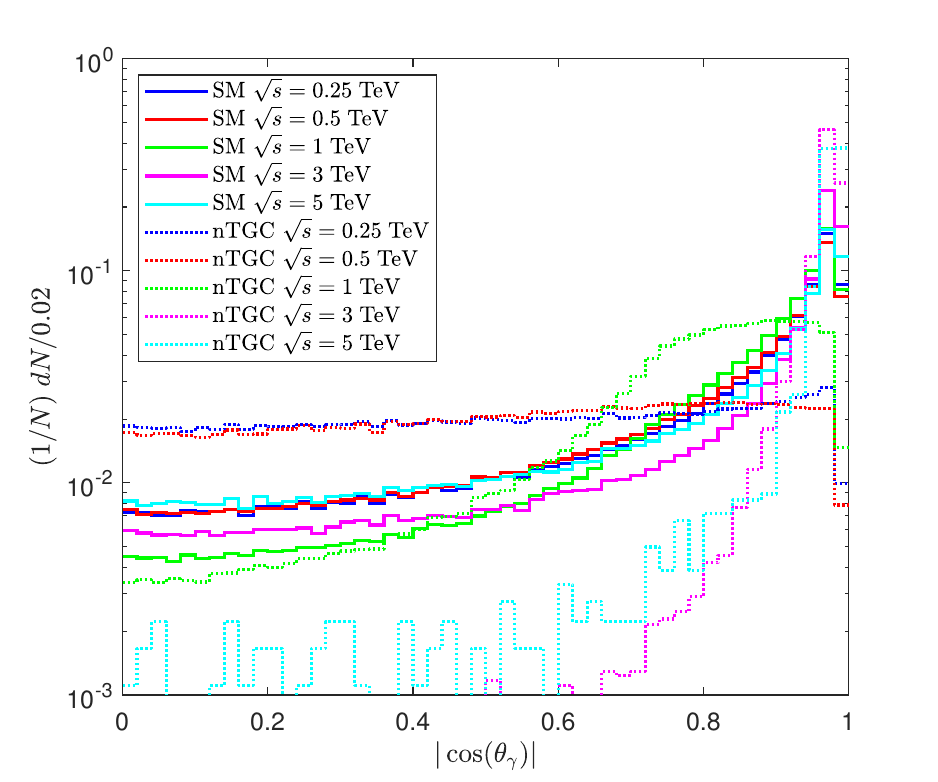}}
\subfigure[]{\includegraphics[width=0.49\textwidth]{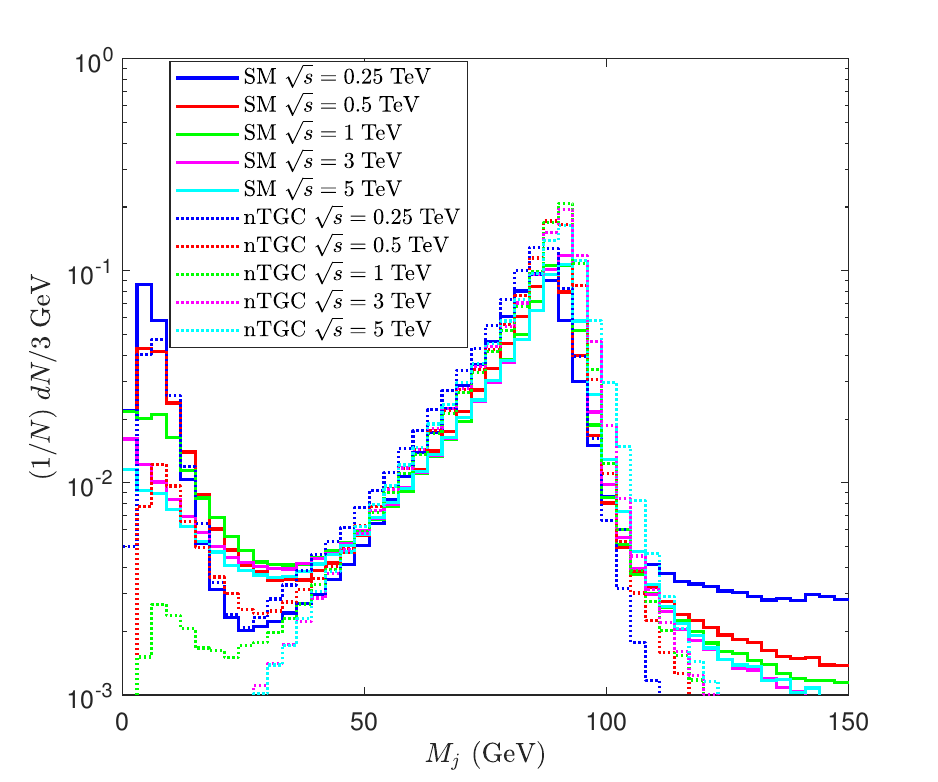}}
\subfigure[]{\includegraphics[width=0.49\textwidth]{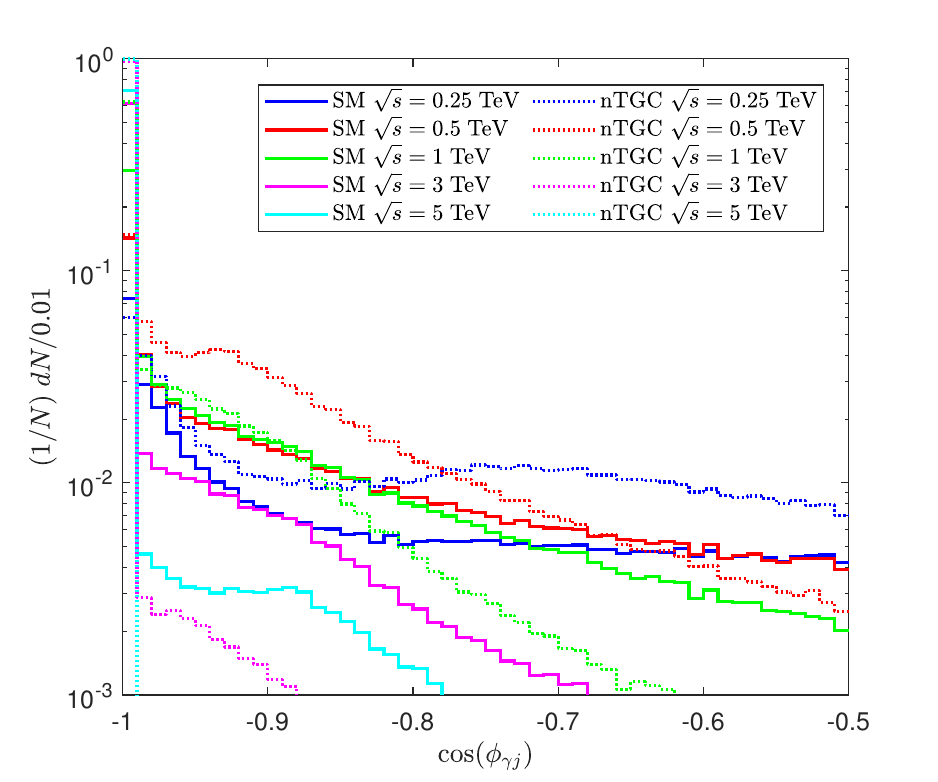}}
\caption{\label{Fig:KinematicFeaturejj}The normalized distributions of $|\cos \theta _{\gamma}|$, $M_j$,  $|\cos \phi _{\gamma j}|$ for hadronic Z decays.}}
\end{figure*}

Similar as the case of leptonic decay, we use $M_j$ and $|\cos (\theta _{\gamma})|$ to discriminate the signals from the background, where $M_j$ is the invariant mass of all jets.
The normalized distributions of $M_j$ and $|\cos (\theta _{\gamma})|$ are shown in Fig.~\ref{Fig:KinematicFeaturejj}.~(a) and (b).
We require the $M_j$ to be in the range of $[M_z-45\;{\rm GeV}, M_z+15\;{\rm GeV}]$, and cut off the events with a large $|\cos (\theta _{\gamma})|$.

To remove the background such that the photon is emitted from a jet, we require a small $\cos (\Delta \phi _{\gamma j})$, where $\Delta \phi _{\gamma j}$ is the smallest absolute difference between the azimuth angles of the photon and the jets.
The normalized distributions of $\cos (\Delta \phi _{\gamma j})$ are shown in Fig.~\ref{Fig:KinematicFeaturejj}.~(c).

\begin{table*}
\begin{center}
\begin{tabular}{c|c|c|c|c|c|c|c|c|c|c}
 & \multicolumn{2}{c|}{250 GeV} & \multicolumn{2}{c|}{500 GeV} & \multicolumn{2}{c|}{1 TeV} & \multicolumn{2}{c|}{3 TeV} & \multicolumn{2}{c}{5 TeV} \\
\hline
 & SM & nTGC & SM & nTGC & SM & nTGC & SM & nTGC & SM & nTGC \\
 \hline
$N_{j,\gamma}$ cut & $3915.7$ &  $3222.8$ & $959.3$ & $2857.8$ & $264.0$ &  $1153.4$ & $34.5$ & $434.4$ & $13.1$ & $538.7$ \\
\hline
$M_z-45<M_j<M_z+15$ GeV & $1869.3$ & $1978.3$ & $445.9$ & $2128.5$ & $126.1$ & $905.8$ &  $16.7$ & $343.1$ & $6.27$ & $419.3$ \\
\hline
$|\cos(\theta _{\gamma})|<0.9$ & $1024.1$ & $1762.1$ & $265.4$ & $1945.7$ & $75.7$ & $816.5$      & & & & \\
\hline
$\cos(\phi _{\gamma j}) < -0.99$ &  &  &  &  &  &  & $13.2$ & $334.3$ & & \\
$\cos(\phi _{\gamma j}) < -0.9995$ &  &  &  &  &  &  &  &  & $5.9$ & $418.0$ \\
\end{tabular}
\end{center}
\caption{\label{Tab:cutflowjj}The event selection strategy and cross-sections~(fb) after cuts for hadronic Z decays.}
\end{table*}

The proposed event selection strategy for the hadronic decay and the cross-sections after cuts are summarised in Table~\ref{Tab:cutflowjj}.
In the region of the coefficients we use, at large energies, the cross-sections of the signals are larger than the SM backgrounds.

\begin{figure*}[!htbp]
\centering{
\subfigure[$\sqrt{s}=250$ GeV]{\includegraphics[width=0.32\textwidth]{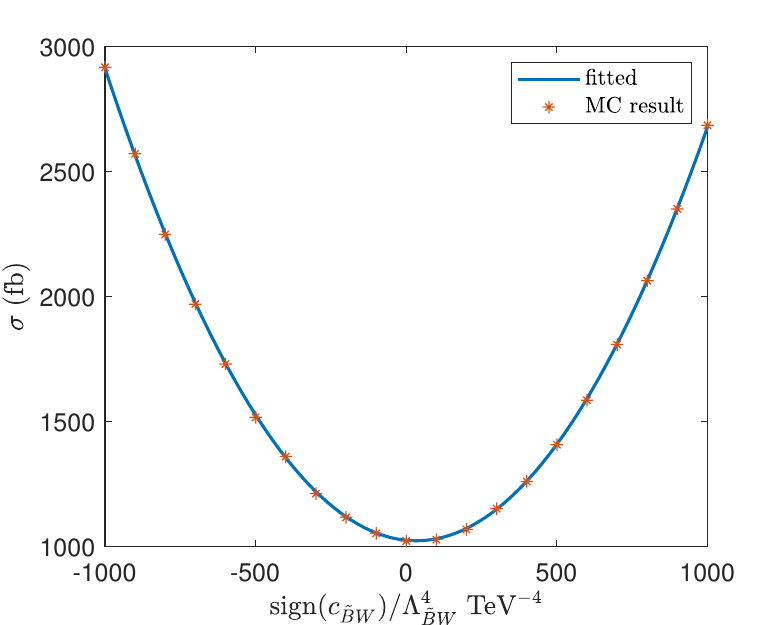}}
\subfigure[$\sqrt{s}=500$ GeV]{\includegraphics[width=0.32\textwidth]{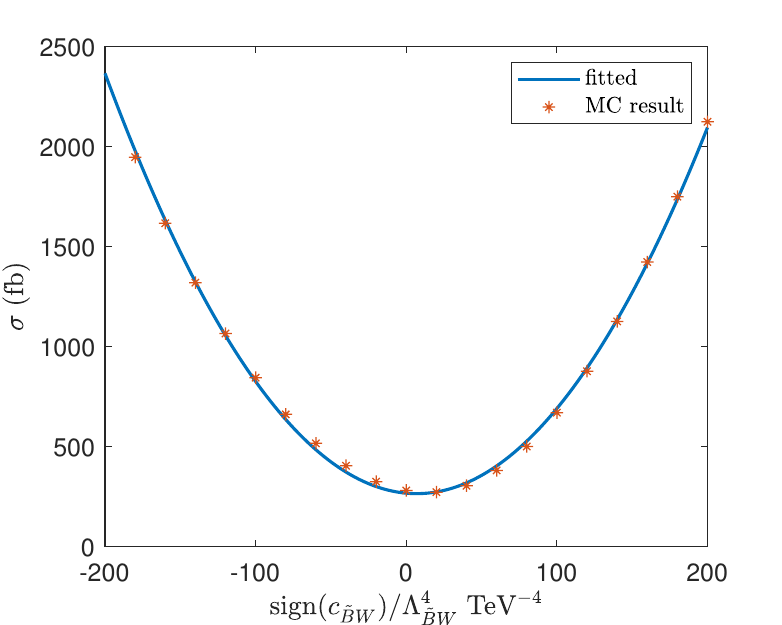}}
\subfigure[$\sqrt{s}=1$ TeV]{\includegraphics[width=0.32\textwidth]{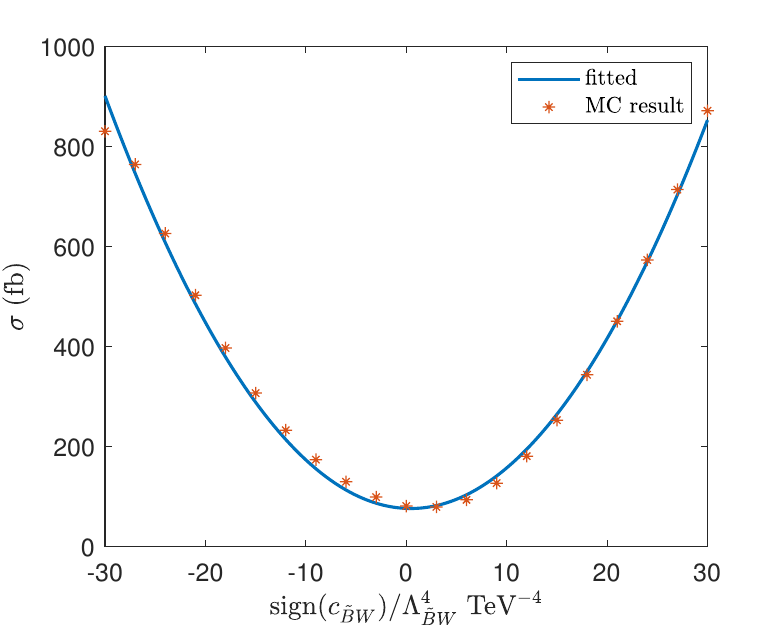}}\\
\subfigure[$\sqrt{s}=3$ TeV]{\includegraphics[width=0.32\textwidth]{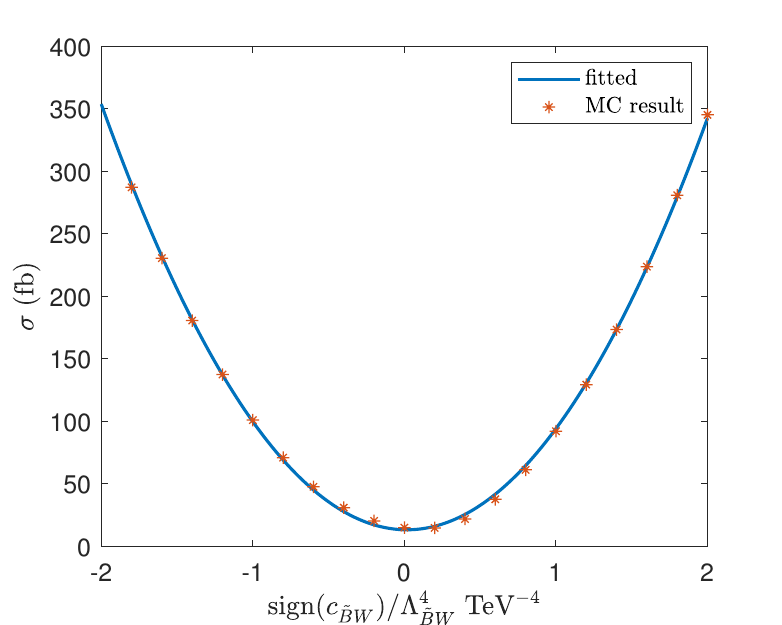}}
\subfigure[$\sqrt{s}=5$ TeV]{\includegraphics[width=0.32\textwidth]{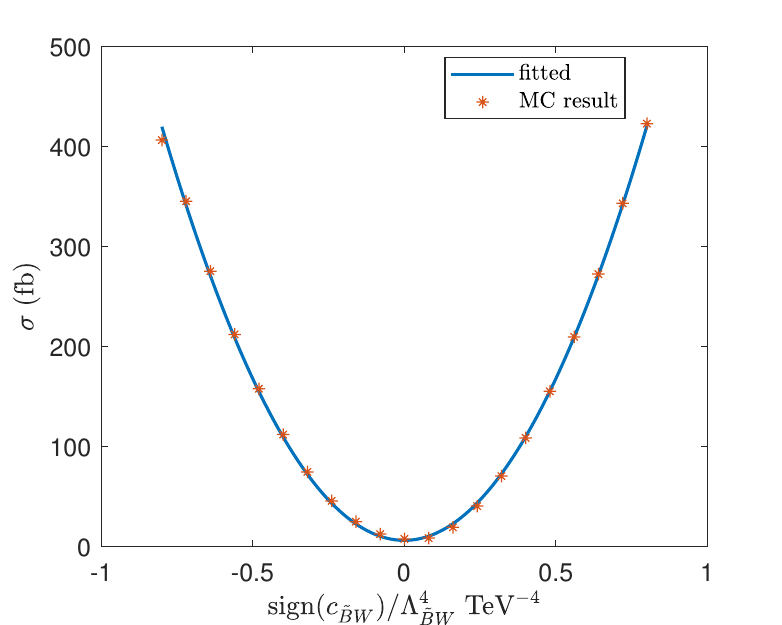}}
\caption{\label{Fig:csjj}Cross-sections as functions of ${\rm sign}(c_{\tilde{B}W})/\Lambda ^4_{\tilde{B}W}$, for hadronic Z decays.}}
\end{figure*}

\begin{figure*}[!htbp]
\centering{
\subfigure[$\sqrt{s}=250$ GeV]{\includegraphics[width=0.32\textwidth]{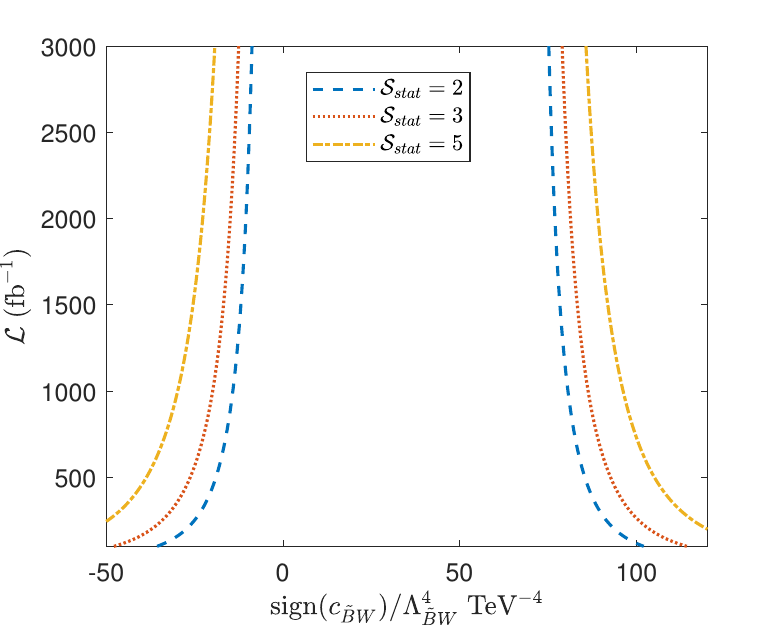}}
\subfigure[$\sqrt{s}=500$ GeV]{\includegraphics[width=0.32\textwidth]{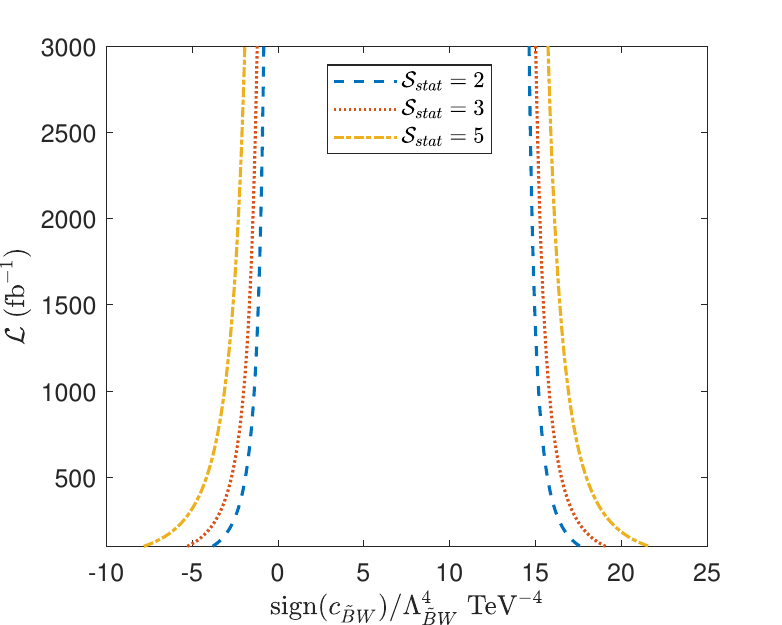}}
\subfigure[$\sqrt{s}=1$ TeV]{\includegraphics[width=0.32\textwidth]{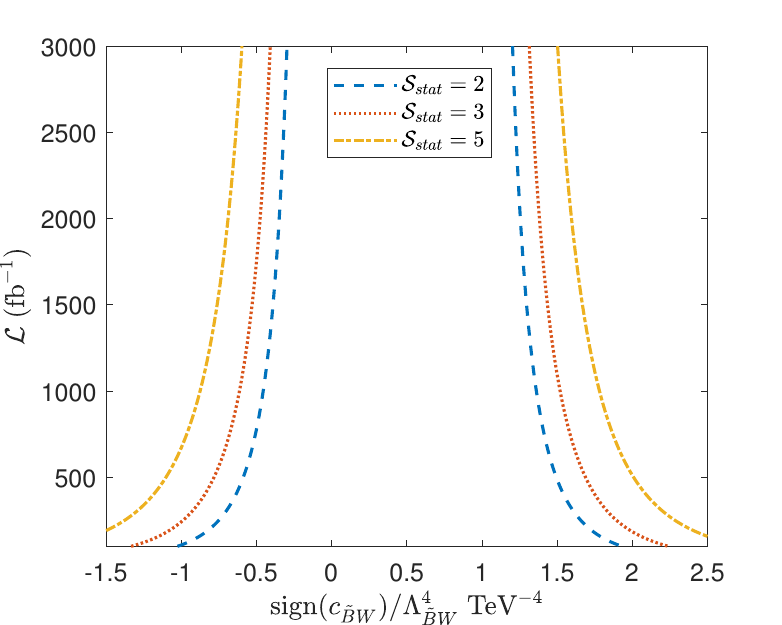}}\\
\subfigure[$\sqrt{s}=3$ TeV]{\includegraphics[width=0.32\textwidth]{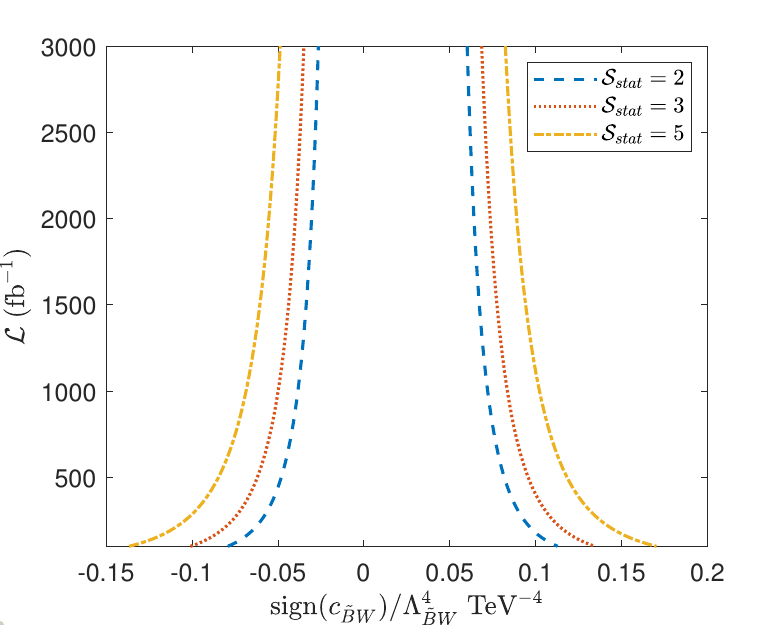}}
\subfigure[$\sqrt{s}=5$ TeV]{\includegraphics[width=0.32\textwidth]{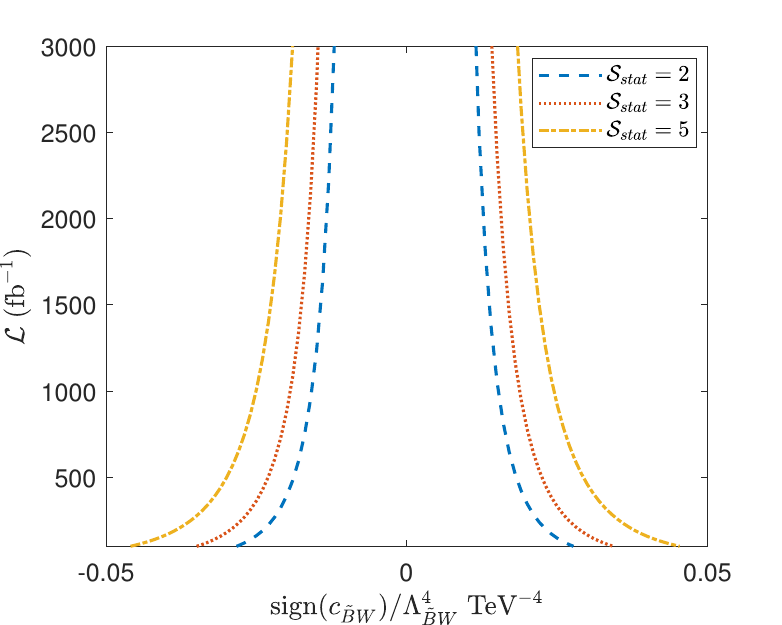}}
\caption{\label{Fig:ssjj}The constraints corresponding to different luminosities for hadronic Z decays. The vertical lines are unitarity bounds.}}
\end{figure*}

\begin{table*}
\begin{center}
\begin{tabular}{c|c|c|c|c}
 $\sqrt{s}=250$ GeV & $\sqrt{s}=500$ GeV & $\sqrt{s}=1$ TeV & $\sqrt{s}=3$ TeV & $\sqrt{s}=5$ TeV\\
\hline
 $0.0005\;{\rm pb}^{-1}$ & $0.0014\;{\rm pb}^{-1}$ & $0.0053\;{\rm pb}^{-1}$ & $0.042\;{\rm pb}^{-1}$ & $0.12\;{\rm pb}^{-1}$ \\
\end{tabular}
\end{center}
\caption{\label{Tab:lumolosityjj}The required luminosities for the constraints obtained by experiments at $\mathcal{S}_{stat}=5$ to be tighter than the unitarity bounds for hadronic Z decays.}
\end{table*}

By scanning the parameter spaces listed in Table~\ref{Tab:region}, the cross-sections of the process $e^+e^-\to jj \gamma$ are obtained and shown in Fig.~\ref{Fig:csjj}.
Compared with the leptonic decay, the signal is more significant for the hadronic decay.
The constraints corresponding to different luminosities~(denoted as $\mathcal{L}$) are shown in Fig.~\ref{Fig:ssjj}.
We do not show the unitarity bounds in the figures because the constraints by unitarity are lenient.
For $\mathcal{S}_{stat}=5$, the required luminosities for the constraints to be tighter than the unitarity bounds are listed in Table~\ref{Tab:lumolosityjj}.
Using $\mathcal{L}=2\;{\rm ab}^{-1}$ as a representation, the expected constraints in experiments are shown in Table~\ref{Tab:constraintsjj}.
Similar as Ref.~\cite{ntgc3}, we find the signal of the $\mathcal{O}_{\tilde{B}W}$ operator is more significant in hadronic $Z$ decays.
Apart from that, the contributions of the interference terms are more important compared with the process $e^+e^-\to \ell^+\ell^-\gamma$.

\begin{table*}
\begin{center}
\begin{tabular}{c|c|c|c|c|c}
 $\mathcal{S}_{stat}$ & $\sqrt{s}=250$ GeV & $\sqrt{s}=500$ GeV & $\sqrt{s}=1$ TeV & $\sqrt{s}=3$ TeV & $\sqrt{s}=5$ TeV\\
\hline
 $2$ & $[-10.5, 76.9]$ & $[-1.0, 14.8]$ & $[-0.35, 1.3]$ & $[-0.030, 0.064]$ & $[-0.013, 0.013]$ \\
 $3$ & $[-14.9, 81.3]$ & $[-1.5, 15.2]$ & $[-0.48, 1.4]$ & $[-0.040, 0.074]$ & $[-0.016, 0.016]$ \\
 $5$ & $[-22.7, 89.1]$ & $[-2.3, 16.1]$ & $[-0.69, 1.6]$ & $[-0.055, 0.089]$ & $[-0.020, 0.020]$ \\
\end{tabular}
\end{center}
\caption{\label{Tab:constraintsjj}The expected constraints on ${\rm sign(c_{\tilde{B}W})}/\Lambda _{\tilde{B}W}^4$ (${\rm TeV}^{-4}$) at $\mathcal{L}=2\;{\rm ab}^{-1}$ for hadronic Z decays.}
\end{table*}

\section{\label{level5}Summary}

The $ZV\gamma$ vertices provide a unique opportunity to study the dimension-8 physics because there is no $ZV\gamma$ vertices in the SM, and there is no dimension-6 operators contributing to the $ZV\gamma$ vertices.
The nTGCs can contribute to the processes $e^+e^-\to \ell^+\ell^-\gamma$ and $e^+e^-\to jj\gamma$ via $ZV\gamma$ vertices.
The contribution of nTGCs to this process can be studied in the $e^+e^-$ colliders such as CEPC.
In this work, we investigate how the above processes are affected by nTGCs.

Whether the EFT is the valid at a large energy scale is an important issue, especially when we need a large energy scale to probe the signals of high dimensional operators.
We study this problem by using partial wave unitarity.
Because the energy scale of the process $e^+e^-\to Z\gamma$ is just the energy of the collider for a $e^+e^-$ collider, the unitarity bounds can be obtained as bounds on the coefficients of operators, which are listed in Table.~\ref{Tab:unitaritybound}.
Since the EFT is invalid when the unitarity bounds are not satisfied, the constraints obtained by the experiments only make sense when the constraints are tighter than the unitarity bounds.
As a consequence, to study the signals of nTGCs, there exists a minimal luminosity given the energy of the collider.
To address this issue, we study the kinematic features of the signal and background events by MC simulation, and the event selection strategy for nTGCs is proposed.
Then the required luminosities to study the nTGCs at different c.m. energies are obtained and presented in Table.~\ref{Tab:lumolosity}.
The sensitivity of the processes $e^+e^-\to\ell^+\ell^-\gamma$ and $e^+e^-\to jj\gamma$ to the nTGCs at future $e^+e^-$ colliders are also studied and the expected constraints are calculated, indicating that these processes are sensitive to the nTGCs.

\section*{ACKNOWLEDGMENT}

\noindent
This work was supported in part by the National Natural Science Foundation of China under Grant Nos. 11905093, 11875157, 12047570 and 11947402, the Natural Science Foundation of the Liaoning Scientific Committee No.2019-BS-154 and the Outstanding Research Cultivation Program of Liaoning Normal University (No.21GDL004).

\bibliography{nTGC1}
\bibliographystyle{h-physrev}

%\end{multicols}

\end{document}